\documentclass[preprint] {article}
\usepackage{amsmath}
\usepackage{epsfig}
\usepackage{amsfonts}
\usepackage{amssymb}
\usepackage{geometry}
\usepackage{graphicx}
\allowdisplaybreaks
\newcommand {\bp}{\begin{pmatrix}}
\newcommand {\ep}{\end{pmatrix}}
\newcommand{\be}{\begin{equation}} \newcommand{\ee}{\end{equation}}
\newcommand{\bea}{\begin{eqnarray}}\newcommand{\eea}{\end{eqnarray}}

\begin{document}
\title{On the bound states and correlation functions of a class of
Calogero-type quantum many-body problems with balanced loss and gain }

\author{Debdeep Sinha\footnote{{\bf email:}  debdeepsinha.rs@visva-bharati.ac.in} 
\ and Pijush K. Ghosh \footnote {{\bf email:}
pijushkanti.ghosh@visva-bharati.ac.in}}
\date{Department of Physics, Siksha-Bhavana, \\ 
Visva-Bharati University, \\
Santiniketan, PIN 731 235, India.}
\maketitle
\begin{abstract}
The quantization of many-body systems with balanced loss and gain is
investigated. Two types of models characterized by either translational
invariance or rotational symmetry under rotation in a pseudo-Euclidean space
are considered. A partial set of integrals of motion are constructed for each
type of model. Specific examples for the translational invariant systems
include Calogero-type many-body systems with balanced loss and gain, where each
particle is interacting with other particles via four-body inverse-square
potential plus pair-wise two-body harmonic terms. A  many-body system
interacting via short range four-body plus six-body inverse square potential
with pair-wise two-body harmonic terms in presence of balanced loss and gain
is also considered. In general, the eigen values of these two models contain
quantized as well as continuous spectra. A completely quantized spectra and
bound states involving all the particles may be obtained by employing
box-normalization on the particles having continuous spectra. The normalization
of the ground state wave functions in appropriate Stoke wedges is discussed.
The exact n-particle correlation functions of these two models are obtained
through a mapping of the relevant integrals to known results in random matrix
theory. It is shown that a rotationally symmetric system with generic
many-body potential does not have entirely real spectra, leading to unstable 
quantum modes. The eigenvalue problem of a Hamiltonian system with
balanced loss and gain and admitting dynamical $O(2,1)$ symmetry is also
considered.
\end{abstract}

{\bf keywords:} Quantum many-body system;  Exactly solvable models;
Calogero-type models; Correlation functions; Dissipative system

\tableofcontents{}
\section{Introduction}

Dissipation is an ubiquitous natural phenomenon. One of the simplest examples
of dissipative system is the damped harmonic oscillator with a friction term
linear in velocity. Investigation on  Hamiltonian formulation of this system
was undertaken by Bateman \cite{bat} more than eighty five years ago. The
Bateman's Hamiltonian is defined in an ambient space with twice the degrees
of freedom of the original oscillator. The extra degree of freedom constitutes
an auxiliary system which is the time-reversed version of the original 
dissipative oscillator. Therefore, the gain due to one oscillator is equally
balanced due to the loss from the other. Neither the dissipative nor the
auxiliary oscillator alone defines a Hamiltonian system, rather it is
the combination of the two oscillators which is described by Bateman
Hamiltonian. A balanced loss and gain system is the one for which the flow
preserves the volume in the  position-velocity state space,
although individual particles may be subjected to gain and/or loss.
Thus, the Bateman oscillator(BO) is an example of Hamiltonian system with
balanced loss and gain. The quantization of BO
was discussed in Ref.\cite{fes} with a motivation to give a macroscopic
description of deep inelastic scattering. Various subtle issues and important
results related to quantization of BO are discussed in the
literature \cite{bopp,trikochinsky, dekker,rasetti, rabin,jur}.

The BO neither admits classically stable solutions nor quantum bound states.
Further, the dynamics of the dissipative and the auxiliary oscillators are
independent of each other. However, the situation changes significantly,
if an appropriate ${\cal{PT}}$ symmetric interaction is added to the Bateman
Hamiltonian that couples the original and the auxiliary oscillators. The
resulting system admits classically stable solutions as well as quantum bound
states for some regions in the parameter space for which the
$\cal{PT}$-symmetry is unbroken \cite{ben}.
An equilibrium is reached for the case of unbroken ${\cal{PT}}$ symmetry so
that the amount of energy gained by one of the oscillators is reverted back to
the other at the equal rate. Similar features have been observed for a large
class of Hamiltonian systems with balanced loss and
gain\cite{ben1,ds-pkg,pkg-ds,ds-pkg6,pkg6}. The experimental results obtained
in Ref. \cite{wgm} in the context of whispering gallery modes are explained
well by the mathematical model of Ref. \cite{ben}. 

The dynamics of the particles subjected to gain and that of loss are
intertwined to each other in the examples considered in Refs.
\cite{ben, ben1,ds-pkg,pkg-ds, ds-pkg6,pkg6}. Further, for the case of
unbroken ${\cal{PT}}$-symmetry, equilibrium states exist for the Hamiltonian.
Consequently, the concept of auxiliary system or ambient space ceases to exist
for these examples and the Hamiltonian of
Refs. \cite{ben, ben1,ds-pkg,pkg-ds,ds-pkg6,pkg6} should not be seen simply as
an extended version of Bateman's oscillator, rather it should be viewed in its
totality as describing a new type of physical system. This indeed is a
paradigm shift in interpreting this class of system with balanced loss and
gain that is
Hamiltonian. As emphasized in Ref. \cite{pkg6}, a system with balanced loss and
gain is not always amenable to a Hamiltonian formulation without the
introduction of an ambient space or auxilliary system. Thus, this new class
of Hamiltonian, where there is no hidden or explicit auxiliary
system\cite{pkg6}, deserves a special attention and teatment from that of
Bateman-type Hamiltonian.

It should be mentioned in this context that another type of
generalization of Bateman's approach is always allowed, where a clear
distinction between original and auxiliary systyems exists.
In particular, Hamiltonian formulation of
dissipative nonlinear system necessitates the introduction of a `unidirectional
coupling' between the system and its auxiliary counterpart\cite{ds-pkg6,sagar}.
The coupling is unidirectional in the sense that the dynamics of the particles
with loss are independent of the dynamics of the particles with gain. However,
the converse is not true and no equilibrium state exists for the Hamiltonian.
This again leads to a Hamiltonian system with balanced loss and gain, albeit
in an ambient space for which a clear distinction between the
system and its auxiliary counterpart exists. Such a formulation is useful
for studying purely dissipative dynamics by using known techniques associated
with a Hamiltonian system like canonical perturbation theory, canonical
quantization, KAM theory etc., which is beyond the scope of the present article.

The emphasis in this article is on many-particle Hamiltonian systems with
balanced loss and gain for which no ambient space or auxiliary system is
introduced. A few exactly solvable systems with balanced loss
and gain of this type are presented along with the construction of a set of
integrals of motion for each model in Ref. \cite{pkg-ds}. Stable
classical solutions are obtained for a class of systems characterized by
translational invariance. However, rotationally invariant systems
under rotation in a pseudo-Euclidean space do not admit any classically
stable solutions. 

The Calogero-Moser-Sutherland (CMS) type of many-body problems\cite{calo,sut}
are  famous examples of exactly solvable models in one dimension. The CMS
models and various generalizations of them have been extensively studied in
the literature\cite{ob,poly,pkg1}. These models have many implications in 
various diverse branches of physics and mathematics such as in exclusion
statistics\cite{exst}, quantum chaos \cite{quch}, spin chains\cite{hal},
algebraic and integrable structure\cite{integ}, self-adjoint
extensions\cite{pkg2}, collective field formulation of many-particle
systems\cite{cfmp}, quantum Hall effect\cite{qhe}, Tomonaga-Luttinger
liquid\cite{tl} etc.. Therefore, an extension of CMS model in the context of
system having balanced loss and gain and the study of their
integrability and/or exact solvability is an obvious curiosity. The rational
CMS model with balanced loss and gain has been analyzed previously for two 
particles and shown to admit stable classical solutions as well quantum bound
states\cite{ds-pkg}. It appears from the analysis in Ref. \cite{pkg-ds} that
the celebrated
CMS systems are not amenable to a Hamiltonian formulation for more than two
particles, if these models are generalized to include balanced loss and gain
terms. However, CMS-type models with balanced loss and gain, where particles
interact with each other through pairwise four-body inverse square plus
two-body harmonic interactions were shown to admit classically stable
solutions. It should be mentioned that the four-body interactions in case of
many-body systems without any gain and/or loss term have been considered
earlier in the literature \cite{wolf,hack,bac}. The exactly solvable quantum
models of CMS-type with translational invariant two and four-body
interactions is investigated in Ref. \cite{hack} and exactly solvable
four-body interaction with translational non-invariant interactions is
discussed in Ref. \cite{bac}. The form of the four-body interactions in
these models are different from those presented in Ref. \cite{pkg-ds}.

The purpose of this article is to consider quantization of the Hamiltonian
system with balanced loss and gain of Ref. \cite{pkg-ds} and investigate
various aspects of these models including eigen spectra. The quantum
Hamiltonian obtained in this case has the form of the Hamiltonian describing
the motion of particles on a pseudo-Euclidean plane and subjected to a uniform magnetic
 field perpendicular to the plane. Therefore, it is always possible to interpret
the Hamiltonian of a balanced loss and gain system as describing the motion of particles
interacting with each other through a potential in the background of a
pseudo-Euclidean metric and is subjected to a uniform “magnetic field” in
specific directions, where the gain/loss parameter plays the role of the
``magnetic field". Further, a unitary transformation may be identified which
plays the role of gauge transformation in relating a Hamiltonian describing
particles in an external uniform magnetic field in symmetric gauge to that
of the same Hamiltonian in the Landau gauge. Two classes of models
characterized by their symmetry properties are considered for detail
investigations: (i) transnational invariant systems and (ii) rotationally
invariant systems under rotation in a pseudo-Euclidean space endowed with
the metric $[g]_{ij}=(-1)^{i+1} \delta_{ij}$. A duality symmetry is
introduced between two types of translational invariant systems and used
to relate the eigenspectra of these models.

The specific examples of translational invariant systems include,
(i) coupled harmonic oscillators, (ii) Calogero-type many-body systems
with balanced loss and gain, where each particle is interacting with
other particles via four-body inverse-square potential plus pair-wise
two-body harmonic terms and (iii) a  many-body system interacting via short
range four-body plus six-body inverse square potential with pair-wise two-body
harmonic terms.  In general, the eigen values of these translational
invariant models contain quantized as well as continuous spectra. The
box-normalization may be used on the particles having continuous spectra
in order to obtain a completely quantized spectra and bound states involving
all the particles. The appropriate Stokes wedges are identified so that
the energy is bounded from below and the ground state wave functions is
normalizable. The exact n-particle correlation functions of the two
Calogero-type models are obtained through a mapping of the relevant integrals
to known results in random matrix theories. 

The Lagrangian of BO is invariant under hyperbolic rotation and the 
corresponding quantum Hamiltonian has the same symmetry. Neither classical
stable states nor quantum bound states are possible for BO. The effect of
a coupling between the two oscillators of Bateman Hamiltonian through
rotationally invariant potential is worth investigating. In this context, a
rotationally symmetric Hamiltonian with generic many-body potential 
respecting this symmetry is investigated in presence of balanced loss
and gain. It is shown that such a system does not admit entirely real spectra
for the generic potential having rotational symmetry. Consequently, there are
quantum modes corresponding to decays and growths and the system is suitable
for describing quantum dissipation.

The BO has a dynamical $SU(1,1)$ symmetry which can be used to completely
solve the eigenvalue problem\cite{rasetti}. The oscillators of Bateman
Hamiltonian may be coupled through specific class of potentials  so that
dynamical $SU(1,1)$ symmetry is maintained.  A many-body system with
balanced loss and gain and admitting dynamical $O(2,1)$ symmetry is
considered. It is shown that the excitations corresponding to radial degree
of freedom can always be obtained analytically for a generic many-body
potential by employing the $O(2,1)$ symmetry. The associated eigenfunctions
are also normalizable in proper Stoke wedges. However, the solutions of the
eigenvalue equation corresponding to angular degrees of freedom depends on
the specific form of the potential.

The plan of the article is as follows. In the next section, the model is
introduced with a  discussion on the quantization of the system. In section
$3$, the exact
solutions, eigen spectra and n-particle correlations are obtained for
translational invariant systems with balanced loss and gain. The
models considered in this section are, (i) coupled harmonic oscillators,
(ii) Calogero-type many-body systems with balanced loss and gain, where each
particle is interacting with other particles via four-body inverse-square
potential plus pair-wise two-body harmonic terms and (iii) a  many-body system
interacting via short range four-body plus six-body inverse square potential
with pair-wise two-body harmonic terms. The eigen-value problems of these
three models are considered in sections 3.1, 3.2.1 and 3.2.3, respectively.
The exact n-particle correlation functions for the systems (ii) and (iii)
are computed in sections 3.2.2 and 3.2.4, respectively. Rotationally invariant
systems are considered in section $4$. Section $5$ deals with systems admitting
$O(2,1)$ symmetry. Finally,  summary and discussions of the results
are presented in section 6.

\section{Quantization of classical Hamiltonian}

A system with balanced loss and gain is the one for which the flow
preserves the volume in the  position-velocity state space,  although individual
particles may be subjected to gain and/or loss. This also implies that all such
systems are closed. In general, many-particle systems with balanced loss and gain
and interacting via an arbitrary potential may not be amenable to Hamiltonian
formulation\cite{pkg6}. A few examples of Hamiltonian systems with balanced loss
and gain were considered in the literature on case-by-case basis in the context of
${\cal{PT}}$ symmetric systems \cite{ben}.  Existence of equilibrium states and
phase-transitions involving broken and unbroken phases of ${\cal{PT}}$ symmetry are
interesting features of these Hamiltonian systems. 
Recently, a Hamiltonian formulation of generic many-body systems having
balanced loss and gain is presented in Ref.\cite{pkg-ds}. Several exactly
solved models, including a class of CMS-type models with four-body
inverse-square and two-body harmonic interactions, have been introduced
with explicit construction of their classical solutions\cite{pkg-ds}. It should be
mentioned here that the Hamiltonian for all these solvable models
is defined on the physical space and there is no ambient space or auxiliary system.
Quantization of these systems with balanced loss and gain will be discussed in this
section. Apart from describing certain general results, eigen spectra of
CMS-type $N$-body problems with balanced loss and gain will be discussed
in some detail.

The general form of a Hamiltonian describing a balanced loss and gain system 
is presented in Ref. \cite{pkg-ds}. For a particular representation of the matrices 
$M, R, D$ as described in Ref. \cite{pkg-ds}, the Hamiltonian describing a balanced 
loss and gain system with $N=2 m$ number of particles may be written in the following form:
\bea
H=\sum_{i=1}^m \left[2p_{2i-1}p_{2i}+\gamma(x_{2i-1}p_{2i-1}-x_{2i}p_{2i})-
\frac{\gamma^2}{2}x_{2i-1}x_{2i}\right]+V(x_i),
\label{Ham}
\eea
\noindent where $\gamma$ is the gain/loss parameter and $V(x_i)$ is the
potential.  
A detail investigation on related issues like how this system represents a balanced loss
and gain and various generalizations of the system described by the Hamiltonian
(\ref{Ham}) may be found in Refs. \cite{pkg-ds, ds-pkg6,pkg6}.
The BO is reproduced\cite{bat} for $m=1$ and
$V=\frac{\omega^2}{2} x_1 x_2$. A generalization of these results to arbitrary $m$ is
straightforward. The Hamiltonian for this case corresponds to $m$ copies of
Bateman oscillators. The quantum problem for the case $m=1$ has been studied in
detail\cite{bopp,trikochinsky, dekker, rasetti,rabin,jur} and no bound states
can be found.  The equations of motion resulting from the $H$
decouple into two differential equations, one with decaying solution and
the other with growing solution. The Hamiltonian system does not admit any
classically stable solutions. However, stable classical solutions are possible
for several  choices\cite{ben,ben1,wgm,pkg-ds} of $V$, including many-particle
Calogero-type models. Quantization of systems admitting classically stable
solutions will be discussed in some detail.
It is known that the Hamiltonian $H$ in Eq. (\ref{Ham}) can always be
reformulated in the background of a pseudo-Euclidean metric\cite{pkg-ds} with
the introduction of a new coordinate system:
\be
z_i^{-}= \frac{1}{\sqrt{2}} \left ( x_{2i-1}-x_{2i} \right ),
\ \ z_i^{+}=\frac{1}{\sqrt{2}} \left ( x_{2i-1}+x_{2i} \right ),
\ i=1, 2,\dots m.
\label{trans}
\ee
\noindent The new co-ordinate system is related to the old one through an
orthogonal transformation with the associated orthogonal matrix given by,
$
\hat{O}=\frac{1}{\sqrt{2}}\left[I_m\otimes\left(\sigma_1+
\sigma_3\right)\right],
$
\noindent where $I_m$ is the $m \times m$ identity matrix and $\sigma_{1},
\sigma_3$ are Pauli matrices.
The Hamiltonian in new coordinates takes the following form:
\be
H=  \sum_{i=1}^m \left [( P_{z_i^+}^2 - P_{z_i^-}^2) +
\gamma \left ( z_i^+ P_{z_i^-} + z_i^- P_{z_i^+} \right ) - 
\frac{\gamma^2}{4} \left \{ (z_i^+)^2 - (z_i^-)^2 \right \} \right ] +
V(z_i^+,z_i^-).
\label{ham-z}
\ee
\noindent The terms linear in $\gamma$ are components of angular momentum 
operators in the pseudo-Euclidean metric $g_{ij}=(-1)^{i+1} \delta_{ij}$.
The Hamiltonian has the standard form in the background of the 
pseudo-Euclidean metric.
 
The canonical quantization of the classical Hamiltonian (\ref{ham-z}) is
achieved by considering the classical variables $P_{z_j^+}, P_{z_j^-}, z_j^+,
z_j^-$ as operators satisfying the commutation relations
\be
 \left [ z_j^+, P_{z_j^+}\right ] = i, \left [ z_j^-, P_{z_j^-} \right ] = i.
\label{commutator}
\ee
\noindent All other commutators involving $P_{z_j^+}, P_{z_j^-}, z_j^+, 
z_j^-$  are taken to be zero. The co-ordinate representation of the operators
$P_{z_j^+}$ and $P_{z_j^-}$ are obtained as follows:
\be
P_{z_j^+}:=-i \partial_{z_j^+}, \ \ P_{z_j^-}:= -i \partial_{z_j^-}.
\label{oper}
\ee 
\noindent With the introduction of the operators,
\be
{\Pi}_{z_i^{\pm}}:= P_{z_i^{\pm}} - A_i^{\pm} =
-i \partial_{z_i^{\pm}} \pm \frac{\gamma}{2} z_i^{\mp}, \\
A_i^{\pm}:= \mp \frac{\gamma}{2} z_i^{\mp},
\ee
the quantum Hamiltonian $\hat{H}$ has the following expression:
\bea
\hat{H} & = & \sum_{j=1}^m\left[\left(P^2_{z_j^+}-
P^2_{z_j^-}\right)+ \gamma\left(z_{j}^{+}P_{z_j^-}+
z_{j}^{-}P_{z_j^+}\right)- \frac{\gamma^2}{4}
\left( ({z_j^+})^2-({z_j^-})^2 \right)\right]+V(z_j^-,z_j^+)
\label{qm11} \\
& = & \sum_{i=1}^m \left [ \left ( {\Pi}_{z_i^+} \right )^2 - 
\left ( {\Pi}_{z_i^-} \right )^2 \right ] + V(z_j^-,z_j^+),
\label{quantumH}
\eea
\noindent where $P_{z_j^-}$, $P_{z_j^+}$ should be understood as operators
defined by Eq.(\ref{oper}).
The operators ${\Pi}_{z_i^{\pm}}$ satisfy the following relations:
\be
\left [ {\Pi}_{z_i^-}, {\Pi}_{z_j^-} \right ] = 0,\ \ \ 
\left [ {\Pi}_{z_i^+}, {\Pi}_{z_j^+} \right ] = 0, \  \
\left [ {\Pi}_{z_i^-}, {\Pi}_{z_j^+} \right ] = - i \gamma \delta_{ij}.
\ee
\noindent It is interesting to note that the Hamiltonian $\hat{H}$  has the
form of a Hamiltonian describing the motion of particles on a pseudo-Euclidean
plane and subjected to a uniform magnetic field along a direction
perpendicular to the plane. 
Therefore, the Hamiltonian $\hat{H}$ can be interpreted as describing a system
of $2m$ particles interacting with each other through the potential
$V(z_i^-,z_j^+)$ in the background of a pseudo-Euclidean metric $g_{ij}=(-1)^{i+1} \delta_{ij}$ 
and is subjected to a uniform ``magnetic field" $\gamma$ in the perpendicular
directions of each plane `$z_i^--z_i^+$'. This analogy is helpful in using
the existing terminologies for systems with magnetic field to decribe different
physical situations described by $\hat{H}$. In particular, the
operators $A_i^{\pm}$  have the form of vector potentials in the symmetric
gauge producing uniform magnetic field for each `$i$'. A unitary transformation
may be used to express $\hat{H}$ in the Landau gauge. In particular,
\bea
&& S:=exp \left [ \frac{i\gamma}{2} \sum_{j=1}^m z_j^+ z_j^- \right ],
\nonumber \\
&& \hat{H}_{L_1} = S^{-1} \hat{H} S = \sum_{i=1}^m \left [
\left ( P_{z_i^+} + \gamma z_i^- \right )^2 - 
P_{z_i^-}^2  \right ] +  V(z_j^-,z_j^+),\nonumber \\
&& \hat{H}_{L_2} = S \hat{H} S^{-1} = \sum_{i=1}^m \left [
P_{z_i^+}^2 - \left (P_{z_i^-} - \gamma z_i^+ \right )^2 
\right ] +  V(z_j^-,z_j^+).
\label{landau}
\eea
\noindent There is no realistic magnetic field in the system and the origin
of gauge degrees of freedom should be explained. In fact, this is related to
the fact that the Lagrangian
corresponding to $\hat{H}$, $\hat{H}_{L_1}$ and $\hat{H}_{L_2}$ differ by
total derivative terms. It may be noted that the concept of analogous
vector potential and associated uniform magnetic field has been discussed
previously\cite{blasone} in the context of BO. The vector potentials
$A_i^{\pm}$ are appropriate generalizations for the many-body interacting
systems with balanced loss and gain. The concept of different gauge choices,
like the symmetric and Landau gauge, in the context of systems with balanced
loss and gain is new, which has not appeared in the literature previously. The
quantum problem is
analyzed in this article for two classes of potentials, (i) translational
invariant potentials and (ii) rotationally invariant potentials. The symmetric
gauge Hamiltonian $\hat{H}$ will be used throughout this article including
rotationally invariant systems, except for section 3, where the systems with translational
invariant Hamiltonian are suitably described in terms of the Landau gauge
Hamiltonian $\hat{H}_{L_1}$ and $\hat{H}_{L_2}$. In fact, the box
normalization for translational invariant systems with $\hat{H}$ is
problematic. The imposition of the periodic
boundary condition on the wave-function of $\hat{H}$ leads to inconsistent
and non-physical results. This is an example of the fact that classical
Lagrangian differing by a total time derivative term may not lead to the same
quantum theory, although the classical dynamics for the two cases are
identical. Thus, the correct route to quantize a translational invariant
classical system with balanced loss and gain is to start with the Lagrangian
that lead to $\hat{H}_{L_1}$ or $\hat{H}_{L_2}$, instead of starting from the
Lagrangian ${\cal{L}}$ corresponding to $\hat{H}$ and following the
sequence ${\cal{L}} \Rightarrow  \hat{H} \Rightarrow \hat{H}_{L_1} (
\hat{H}_{L_2})$.

The parity ${\cal{P}}$ and time-reversal symmetry ${\cal{T}}$ are 
defined as,
\bea
&& {\cal{P}} :z_i^+ \rightarrow - z_i^+, \ z_i^- \rightarrow z_i^-, \
P_{z_i^+} \rightarrow - P_{z_i^+}, \ P_{z_i^-} \rightarrow P_{z_i^-},\nonumber \\
&& {\cal{T}}: z_i^+ \rightarrow z_i^+, \ z_i^- \rightarrow z_i^-, \
P_{z_i^+} \rightarrow -P_{z_i^+}, \ P_{z_i^-} \rightarrow -P_{z_i^-}.
\label{pt1}
\eea
\noindent The parity transformation defined by Eq. (\ref{pt1}) has a proper
geometrical interpretation provided the Hamiltonian $\hat{H}$ is identified as
describing a system of $m$ particles in two dimensions characterized by
$z_i^-, z_i^+$ coordinates. The terms quadratic in the canonical momenta
$P_{z_i^{\pm}}$ in
$\hat{H}$ are invariant under the parity ${\cal{P}}$ and the time-reversal
symmetry ${\cal{T}}$, separately. The same is true for the terms quadratic
in the co-ordinates $z_i^{\pm}$. However, ${\cal{P}}$ and ${\cal{T}}$ symmetries
are broken individually for each term linear in $\gamma$ with the effect of
a change in overall minus sign. Thus, the terms linear in $\gamma$ are
invariant under the combined ${\cal{PT}}$-transformation. The operators
$\Pi_{z_i^{\pm}} \rightarrow \pm \Pi_{z_i^{\pm}}$ under ${\cal{PT}}$
transformation.  The potential $V$ being
real is invariant under ${\cal{T}}$ transformation for its generic form. The
${\cal{PT}}$-symmetry of the Hamiltonian $H$ thus reduces to
${\cal{P}}$-symmetry of the potential. The potential is ${\cal{P}}$-symmetric
provided the following condition is satisfied: 
\be
V(z_i^-, z_i^+) = V(z_i^-, - z_i^+).
\label{condipt}
\ee
\noindent It is interesting to note that for $V=V(z_i^-)$, Eq. (\ref{condipt})
is automatically satisfied and the potential admits a translational symmetry
for translation along all $z^+_i$ directions. The potentials
considered in the next section are of this type and therefore allow both 
translation and parity-time reversal symmetries. It should be emphazied that
there is no relation between the translation symmetry and the
${\cal{P}}$-symmetry, rather the condition (\ref{condipt}) is automatically
satisfied by translational invariant potential. The similarity operator $S$ is
invariant under the combined ${\cal{PT}}$ transformation, leading to the
relations,
\bea
[\hat{H},{\cal{PT}}]=0 \Rightarrow [\hat{H}_{L_1},{\cal{PT}}]=0,\ \  
[\hat{H}_{L_2},{\cal{PT}}]=0,
\eea
\noindent which implies invariance of  $\hat{H}_{L_1}$ and $\hat{H}_{L_2}$
under the same transformation.

The parity transformation is not unique and a second choice of
parity transformation ${\cal{P}}_1$  is as follows:
\be
{\cal{P}}_1 :z_i^+ \rightarrow z_i^+, \ z_i^- \rightarrow - z_i^-, \
P_{z_i^+} \rightarrow P_{z_i^+}, \ P_{z_i^-} \rightarrow - P_{z_i^-}.
\ee
\noindent 
The Hamiltonian $\hat{H}$ or equivalently $\hat{H}_{L_1}$ $(\hat{H}_{L_2})$ are ${\cal{P}}_1
{\cal{T}}$ invariant provided,
\be V(z_i^-, z_i^+) = V(-z_i^-, z_i^+).
\label{condi_p1t}
\ee
\noindent  For $V=V(z_i^+)$, Eq. (\ref{condi_p1t}) is automatically satisfied
and the potential admits a translational symmetry for translation along
all $z^-_i$ directions. The operator $S$ is ${\cal{P}}_1{\cal{T}}$
invariant, while ${\cal{P}}_1{\cal{T}}: \Pi_{z_i^{\pm}} \rightarrow -
\Pi_{z_i^{\pm}}$.
The Hamiltonian is invariant under both ${\cal{PT}}$ as well as 
${\cal{P}}_1{\cal{T}}$ provided $V$ is an even function of its arguments,
\bea
V(z_i^-, z_i^+) = V(-z_i^-, - z_i^+).
\label{condi_p2t}
\eea
It may be noted that $\hat{H}$ can also be interpreted as describing a system
of $2m$ particles in one dimension or a single particle in $N=2m$ dimensions.
The parity transformation for the former case is unique, 
${\cal{P}}_0 : z_i^{\pm} \rightarrow - z_i^{\pm}, P_{z_i^{\pm}} \rightarrow
- P_{z_i^\pm}$, and $\hat{H}$ is not invariant under ${\cal{P}}_0 \cal{T}$.
The parity transformation ${\cal{P}}_{N}$ in $N=2m$ dimensions may be defined
as, ${\cal{P}}_{N} : Z \rightarrow W Z, P_Z \rightarrow W P_Z$, where $W$ is a
$2m \times 2m$ orthogonal matrix with determinant $-1$ and
$Z^T \equiv ( z_1^+, \dots, z_m^+, z_1^-, \dots, z_m^- ),
P_Z^T \equiv ( P_{z_1^+}, \dots, P_{z_m^+}, P_{z_1^-}, \dots, P_{z_m^-} )$.
The Hamiltonian $\hat{H}$ can be rewritten as,
\be
\hat{H} = P_{Z}^T G P_{Z} + \gamma  Z^T R P_{Z}  -\frac{\gamma^2}{4} Z^T G Z + V(z_i^+,z_i^-),
\ee
\noindent where $G:=\sigma_3 \otimes I_m, R:=\sigma_1\otimes I_m$. $\hat{H}$
is ${\cal{P}}_{N}{\cal{T}}$ invariant provided,
\be
\left [ W, G \right ]=0, \ \ \{ W, R \} = 0, \ \
\left [ V, {\cal{P}}_{N}{\cal{T}}\right]=0.
\label{sol-parity}
\ee
\noindent For odd $m$,
the choice $W= G$ allows to identify ${\cal{P}}_{N} = {\cal{P}}_1$, while
$W=-G$ gives ${\cal{P}}_{N} = {\cal{P}}$. A more general choice for odd $m$
is $W:= \sigma_3 \otimes Q$, where $Q$ is an $m \times m$ orthogonal matrix
with unit determinant and can be represented in terms of $m$ dimensional
rotation matrices. For even $m$, it appears that an orthogonal matrix $W$ with
determinant $-1$ satisfying (\ref{sol-parity}) can not be found. Thus, 
${\cal{P}}_N {\cal{T}}$-invariant Hamiltonian $\hat{H}$ is not possible in
$N=4m$ dimensions. It appears that the interpretation of $\hat{H}$ as
describing a system of $m$ particles in two dimensions is more appropriate
than the others from the viewpoint of ${\cal{PT}}$ symmetry. It is known
that classical stable solutions and quantum bound states are obtained in
systems with balanced loss and gain in the unbroken ${\cal{PT}}$ regime
\cite{ben,ben1,ds-pkg,pkg-ds}. In this article, only ${\cal{PT}}$ symmetric 
many-body systems with balanced loss and gain will be considered.

Few comments are in order before the end of this section. The Schwinger-Keldysh
formalism is being used in the context of non-equilibrium many-body systems
in a variety of contemporary topics like driven open quantum systems\cite{diehl,
diehl1,jsm}, time-dependent density-functional theory\cite{leewuen},
relativistic hydrodynamics, physics of black-holes, dynamics of entanglement
in quantum field theory etc.\cite{mr}. It has been shown recently that
the Schwinger-Keldysh action at thermodynamic equilibrium is invariant
under time-reversal plus time-translation transformations\cite{diehl1,jsm,abc}.
This result is worth comparing with that of the systems with balanced
loss and gain, where an equilibrium is reached in regard to energy transfer
between the system and the bath for unbroken ${\cal{PT}}$
symmetry\cite{ben,ben1,ds-pkg}. The invariance under time-translation generates
unitary time-evolution and the symmetry is present for both Schwinger-Keldysh
action at thermodynamic equilibrium as well as for $\hat{H}$ and similar 
models\cite{ben,ben1,ds-pkg}. However, the Schwinger-Keldysh action at
thermodynamic equilibrium is invariant under time-reversal symmetry, whereas
unbroken ${\cal{PT}}$ symmetry is essential for the existence of quantum bound
states\cite{ben,ben1,ds-pkg}. It is shown within this context that the
${\cal{PT}}$ symmetry of a many-body two dimensional system embedded in three
dimensions may also be identified as a non-conventional time-reversal symmetry
$\hat{\cal{T}}_{\epsilon}$ in three dimensions\cite{F.Haake,pkg-ds}:
\be
\hat{\cal{T}}_{\epsilon}= exp(i \epsilon \pi \sum_{i=1}^m J_{z_i^-})
\ {\cal{T}}, \ \ \epsilon=\pm 1,
\label{noncon}
\ee
\noindent where $J_{z_i^-}$ denotes generator of rotation for the $i$-th
particle around $z_i^-$-axis. The canonical transformations generated by
${\cal{PT}} ({\cal{P}}_1{\cal{T}})$ and $\hat{\cal{T}}_{1}(\hat{\cal{T}}_{-1})$
on $z_{i}^{\pm}$ and $P_{z_i^{\pm}}$ are identical. The standard argument that
$[h,{\cal{PT}}]=0$ implies that Hamiltonian $h$ admits entirely real spectra
for unbroken ${\cal{PT}}$ symmetry is also valid if ${\cal{PT}}$ is substituted
with $\hat{\cal{T}}_{\epsilon}$. Further, nontrivial non-hermitian potentials
invariant under $\hat{\cal{T}}_{\epsilon}$ may be chosen as in the case of
systems with ${\cal{PT}}$ symmetry. Thus, it is a matter of choice to identify
the symmetry as ${\cal{PT}}$ or $\hat{\cal{T}}_{\epsilon}$. In this article,
the notion of ${\cal{PT}}$ symmetry will be followed.

In quantum mechanics, the conventional time reversal symmetry may be
replaced with an anti-unitary symmetry in order to describe a realistic system.
For example, a hydrogen atom in a constant magnetic field is not invariant
under the conventional time-reversal symmetry. However, the same
Hamiltonian is invariant under a non-conventional time-reversal symmetry \cite{F.Haake}.
The operator $\hat{\cal{T}}_{\epsilon}$ in Eq. (\ref{noncon}) defines such a 
non-conventional anti-unitary symmetry for the systems considered in
this article. 
 
\section{Translational Invariant Hamiltonian}

In this section, translational invariant systems are considered.
The Hamiltonian $\hat{H}_{L_1}$ in Eq. (\ref{landau}) has translational
invariance for $V_{L_1} \equiv V(\{z_i^-\})$ under the transformation
$L_1: z_i^+ \rightarrow z_i^+ + \eta_i,  z_i^- \rightarrow z_i^-$, where
$\eta_i$'s are $m$ arbitrary constants. A translation of each co-ordinate
 $x_{2i-1} \rightarrow x_{2i-1} + \frac{\eta_i}{2},
x_{2i} \rightarrow x_{2i} + \frac{\eta_i}{2}$ generates the transformation $L_1$.
The conserved quantities ${P}_{z_i^+}$ corresponding to this
symmetry are in involution,
\bea
[\hat{H}_{L_1}, P_{z_i^+}]=0,\ [{P}_{z_i^+},{P}_{z_j^+}]=0.
\ \ \forall \ i,j,
\label{iom1}
\eea
\noindent and may be identified as $m$ integrals of motion. This implies that
the system having $N=2m$ numbers of degree of freedom is at least partially
integrable. In order to explore possible complete integrability of
$\hat{H}_{L_1}$, explicit form of the potential $V_{L_1}$ needs to be
specified. Note that  $V_{L_1}$ and hence, $\hat{H}_{L_1}$ is ${\cal{PT}}$
symmetric due to the condition (\ref{condipt}).

The Hamiltonian $\hat{H}_{L_2}$ in Eq. (\ref{landau}) has translational
invariance for $V_{L_2} \equiv V(\{z_i^+\})$ under the transformation
$L_2: z_i^+ \rightarrow z_i^+,  z_i^- \rightarrow z_i^- + \eta_i$.
The transformations $x_{2i-1} \rightarrow x_{2i-1} + \frac{\eta_i}{2},
x_{2i} \rightarrow x_{2i} - \frac{\eta_i}{2}$ generates the transformation
$L_2$. As in the case of ${\hat{H}}_{L_1}$, the existence of $m$ integrals of
motions ${P}_{z_i^-}$ implies $\hat{H}_{L_2}$ is at least partially integrable.
The potential $V_{L_2}$ and $\hat{H}_{L_2}$ are ${\cal{P}}_1{\cal{T}}$
symmetric due to the condition (\ref{condi_p1t}).
There exists a duality transformation between translational invariant
$\hat{H}_{L_1}$ and $\hat{H}_{L_2}$. In particular,
\be
z_i^- \leftrightarrow z_i^+, \ P_{z_i^-} \leftrightarrow P_{z_i^+}, \
\gamma \leftrightarrow -\gamma, \ {V}_{L_{1}} \leftrightarrow
- {V}_{L_{2}} \Rightarrow \hat{H}_{L_1} \leftrightarrow - \hat{H}_{L_2}.
\label{duality}
\ee
\noindent The duality transformation may be used to relate the eigenspectra
of $\hat{H}_{L_1}$ with $\hat{H}_{L_2}$ and vice verse. In particular, if
$\chi(\{ z_k^-,z_k^+\})$ is an eigenstate of $\hat{H}_{L_1}$ with eigenvalues
$E_{L_1}$, then $H_{L_2}$ has the eigenvalues $-E_{L_1}$ and the eigenstates
$\chi(\{ z_k^+,z_k^-\})$. Although the functional forms of the potentials
${V}_{L_{1}}$ and ${V}_{L_{2}}$ are identical, they correspond to entirely
different physical systems, when expressed in terms of the original coordinates
$x_i$. In fact, $\hat{H}_{L_1}$ and $\hat{H}_{L_2}$ are not related to each
other through any similarity transformation for non-vanishing $V_{L_1}$ and
$V_{L_2}$. In this article, the eigenvalue problem of $\hat{H}_{L_1}$ will
be studied in some detail. The eigenvalues and eigenspectra of  $\hat{H}_{L_2}$
will be determined by using  the duality symmetry.

It is always possible to choose a basis such that simultaneous eigen states of
$\hat{H}_{L_1}$ and ${P}_{z_i^+}$ are constructed, since $\hat{H}_{L_1}$ and
${P}_{z_i^+}$ commute. The eigen function of the operator 
${P}_{z_j^+}$ having the continuous eigen value $k_j$, is of the
following form:
\bea
\chi(\{ z_k^-,z_k^+\}) = 
\psi(\{z_k^-\}) \exp \left [ {i}\sum_{j=1}^m z_j^+k_j \right ],
\label{eigpi}
\eea
\noindent where $\psi(\{z_k^-\})$ is a function of the co-ordinates $z_k^-$
only. Substituting (\ref{eigpi}) in the time independent Schrodinger equation,
\bea
\hat{H}_{L_1} \chi=E\chi,
\label{tisc}
\eea
the following equation is obtained:
\bea
\sum_{j=1}^m \left [\partial^2_{z_j^-}
+ \gamma^2 \left ( z_j^- + \frac{k_j}{\gamma} \right )^2 \right ] \psi+
V_{L_1}(z_j^-) \psi = E \psi.
\label{iso}
\eea
\noindent The original eigenvalue problem is defined in terms of $2m$ 
independent co-ordinates $(z_i^-, z_i^+)$. However, the Eq. (\ref{iso})
involves only $m$ co-ordinates $z_i^-$. The decoupling of $m$ co-ordinates
$z_i^+$ from the eigenvalue equation is due to the judicious choice of the
wave-function $\chi$ in Eq. (\ref{eigpi}). 
With the introduction of a new co-ordinate,
\be
\tilde{z}_j = z_j^- + \frac{ k_j}{\gamma},
\ee
\noindent the  Eq. (\ref{iso}) can be rewritten
as an eigenvalue equation in terms of an effective Hamiltonian $H_{eff}$ and
energy $E_{eff}$. In particular,
\be
H_{eff} \psi = E_{eff} \psi, 
\label{effeqn}
\ee
\noindent where,
\be
H_{eff} = - \sum_{j=1}^m \partial^2_{\tilde{z}_j} + V_{eff},\ \
V_{eff}= - \gamma^2 \sum_{j=1}^m\tilde{z}_j^2 - V_{L_1}(\tilde{z}_j),
E_{eff}=-E.
\label{iso12}
\ee
\noindent Thus, if the eigenvalue equation (\ref{effeqn}) in terms of the
effective Hamiltonian with $m$ degrees of freedom is exactly solved for
specific choices of $V_{L_1}$, the same is true for the Hamiltonian $H_{L_1}$ with
$2m$ degrees of freedom. It should be mentioned here that for $V_{L_1}=0$, the
Eq. (\ref{iso12}) decouples into $m$ second order differential equations, where
each equation describes eigen value equation of an ``inverted harmonic
oscillator" with shifted origin. No bound states are possible for this case.
However, proper choices of $V_{L_1}$ can convert the ``inverted harmonic oscillator"
to simple harmonic oscillator plus some desirable interactions. A few such
examples are discussed below.

\subsection{Simple Harmonic Oscillator}

For the choice of the potential,
\be
V_{{L}_1}(z_i^-) = -\frac{\omega^2}{2} \sum_{i=1}^m (z^-_i)^2,
\label{coup}
\ee
\noindent  and $\gamma^2 \neq \omega^2$, Eq. (\ref{iso}) can be re-written as,
\bea
&& \sum_{j=1}^m\left[ - \partial^2_{z_j}
+\Omega^2 z_j^2 \right ] \psi = (-\tilde{E}) \psi,\nonumber \\
&&  \Omega^2=\frac{1}{2}(\omega^2-2\gamma^2), \ \
\tilde{E}=E-\sum_{j=1}^m\frac{\omega^2 k_j^2}{2 \Omega^2},
\ \ z_j= z_j^- -  \frac{\gamma k_j}{\Omega^2}.
\label{iso13}
\eea
\noindent The modified angular frequency $\Omega=\pm \sqrt{\frac{\omega^2-
2 \gamma^2}{2}}$ can take positive as well as negative values. As will be seen
later, the normalization of the wave-function requires $\Omega < 0$. 
This is related to the fact that the eigen-value equation of a simple
harmonic oscillator is different from that of first equation in 
Eq. (\ref{iso13}) due to the appearance of a negative sign in the right hand
side of this equation. The reality of $\Omega$ is ensured for
$-\frac{\omega}{\sqrt{2}} \leq \gamma \leq \frac{\omega}{\sqrt{2}}$. 
For $m=1$, the Hamiltonian $H$ actually describes a Bateman oscillator
with a coupling term of the form $-\frac{\omega^2}{4} (x_1^2+x_2^2)=-
\frac{\omega^2}{4} [ (z_1^-)^2 + (z_1^+)^2]$, which has been considered
previously\cite{ben}. In fact, a choice of the parameter $\epsilon=-\omega^2$
in Eq. (4) of Ref. \cite{ben} reduces to the model under consideration in this
section. The choice of the parameter corresponds to broken ${\cal{PT}}$
regime and it should be mentioned here that the specific type of solutions
presented for this model in this article have not been discussed previously
in the literature. Further, the normalization of wave-functions for both the
long-range and short-range Calogero-type models with confining harmonic term
is similar to the case of pure harmonic oscillators. Thus, the present
example will also be used to explain the normalization scheme for Caloger-type
models to be discussed in next sections.

The first equation of Eq. (\ref{iso13}) is separable into eigenvalue
equations of $m$ oscillators with frequency $\Omega$. The eigenvalues and
eigenfunctions are,
\bea
&& E_{\{n_i\},\{k_i\}} = \sum_{i=1}^m \left [ - \left ( n_i + 
\frac{1}{2} \right )2 \Omega +
\frac{\omega^2k_i^2}{2\Omega^2} \right ], \ \{n_i\} \in 
\mathbb{N}^*, \ \{k_i\} \in \Re, \nonumber \\ 
&& \chi_{\{n_i\},\{k_i\}} = \left ( \frac{\Omega}{\pi}\right)^{\frac{m}{4}}
\prod_{i=1}^m \frac{1}{\sqrt{2^{n_i}n_i!}} H_{n_i}(\sqrt{\Omega} z_i)
\exp \left [-\frac{\Omega}{2} \sum_{i=1}^m z_i^2 +i
\sum_{j=1}^m z_j^+k_j \right ],
\label{eexp}
\eea
\noindent where $H_{n_i}(\sqrt{\Omega} z_i)$ is the Hermite polynomial.
The energy $E_{\{n_i\},\{k_i\}}$ is bounded from below and positive definite
for $ \Omega < 0$. The asymptotic nature of $\chi$ is given by
\bea
\chi \sim \exp[-\frac{\Omega}{2} \sum_{i=1}^m z_i^2 +
i\sum_{j=1}^m z_j^+k_j ]
\label{asymp}
\eea 
\noindent and is not  normalizable along  the real $z_i$ lines for 
$\Omega <0$.  It is required  to fix the proper Stoke wedges
in order to have normalizable solutions. 
\noindent For $\Omega<0$, in Eq. (\ref{asymp}) the first term in 
the exponential becomes $\frac{\Omega}{2}\sum^m_{i=1}z_i^2$.
An extension of the $z_i, \ \forall \ i$, in complex plane gives 
$\sum_{i=1}^m z_i^2=\sum_{i=1}^m\cos(2\theta_i)+i\sin(2\theta_i)$.
For normalizable solution the real part of $\sum^m_{i=1}z_i^2$ 
should be negative, i.e.
\bea
\sum_{i=1}^m\cos(2\theta_i)<0. 
\label{stokewedge}
\eea
A possible solution of which may be obtained for $\theta_i=
\theta,\ \forall \ i$. In this case, $Re(\sum_{i=1}^m  z_i^2)=
m\cos(2\theta)$. Therefore, the exponential part containing 
$\sum_{i=1}^m z_i^2$ vanishes in a pair of Stoke wedges with 
opening angle $\frac{\pi}{2}$ and centered about the positive
and negative imaginary axes in the complex $z_i$-planes. 
It should be mentioned that the Stoke wedge for which the wave function is normalizable is
 not unique. Any possible solution satisfying condition (\ref{stokewedge}) 
gives a Stoke wedge in which the wave function is normalizable. For example, for $m=2$, 
one needs to have $\cos(2\theta_1)+\cos(2\theta_2)<0$ in order to get a normalizable solution. 
One possibility of achieving this is to take both $\cos(2\theta_1)$ and $\cos(2\theta_2)$ negative. 
In this case the Stoke wedge for each complex $z_i$ planes are 
as discussed above. Another possibility is to take $\cos(2\theta_1)$ negative and 
$\cos(2\theta_2)$ positive but $|\cos(2\theta_1)|>|\cos(2\theta_2)|$.  Now, $\cos(2\theta_1)$
 is negative for $\theta_1$ in the ranges $\frac{\pi}{4}<\theta_1<\frac{3\pi}{4}$ and $\frac{5\pi} 
{4}<\theta_1<\frac{7\pi}{4}$ and $\cos(2\theta_2)$ is positive for $\theta_2$ 
in the ranges $\frac{-\pi}{4}<\theta_2<\frac{\pi}{4}$ and
 $\frac{-5\pi}{4}<\theta_2<\frac{5\pi}{4}$. In order to have normalizable solutions one may take
values of $\theta_1$ and $\theta_2$ from the specified ranges such that condition (\ref{stokewedge})
is satisfied.

Another issue of normalization arises because of the plane-wave
nature of the wave-function in $z_i^+$ directions.
The system is not normalizable unless it is bounded along the $z_i^+,
\ \forall \ i$ directions. In order to address this issue the system is assumed 
to be bounded within the length $L$ along $z_i^+, \ \forall \ i$ directions 
with a periodic boundary condition $\chi(z_i^++L,z_i)=\chi(z_i^+,z_i)$.
This boundary condition necessitates the quantization of $k_j=\frac{2 \pi}{L}
l_j, l_j=0,\pm 1, \pm 2, \dots$. The energy  eigenvalues in Eq. (\ref{eexp})
contain discrete as well as continuous spectra. After the box normalization,
the energy gets completely quantized,
\be
E_{\{n_i\},\{m_i\}} = \sum_{i=1}^m \left [ - \left ( n_i +
\frac{1}{2} \right )2 \Omega +
\frac{2 \pi^2 \omega^2 l_i^2}{L^2 \Omega^2} \right ], \ \{n_i, l_i\} \in
\mathbb{N}^*, \ \Omega < 0.
\ee
\noindent The same normalization scheme, i.e. (i) fixing the Stoke wedge and
(ii) employing box normalization along $z_i^+$ coordinates, will be used for
the CMS-type long-range as well as short-range models considered
in section 3.2.

The eigenvalue problem of $\hat{H}_{L_2}$ with the potential
\be
V_{{L}_2}(z_i^+) = \frac{\omega^2}{2} \sum_{i=1}^m (z^+_i)^2,
\ee
\noindent can be studied by using the duality transformation (\ref{duality}).
The potential corresponds to the choice of the parameter $\epsilon=\omega^2$
in Eq. (4) of Ref. \cite{ben} and the system is in broken ${\cal{PT}}$
regime. The energy eigenvalues and eigenfunctions of $\hat{H}_{L_2}$ are
found using Eq. (\ref{duality}) as,
\bea
&& E_{\{n_i\},\{k_i\}} = \sum_{i=1}^m \left [ \left ( n_i +
\frac{1}{2} \right )2 \Omega -
\frac{\omega^2k_i^2}{2\Omega^2} \right ], \ \{n_i\} \in
\mathbb{N}^*, \ \{k_i\} \in \Re, \nonumber \\
&& \chi_{\{n_i\},\{k_i\}} = \left ( \frac{\Omega}{\pi}\right)^{\frac{m}{4}}
\prod_{i=1}^m \frac{1}{\sqrt{2^{n_i}n_i!}} H_{n_i}(\sqrt{\Omega} \tilde{z}_i^+)
\exp \left [-\frac{\Omega}{2} \sum_{i=1}^m (\tilde{z}_i^+)2 +i
\sum_{j=1}^m z_j^-k_j \right ],
\eea
\noindent with $\tilde{z}_i^+=z_i^++\frac{\gamma k_j}{\Omega^2}$.
The wave-function is normalizable along the real axis for
$\Omega >0$. However, the eigenvalue is neither bounded from below nor
from above. On the other hand, the eigenvalue is bounded from above for
$\Omega < 0$ and the wave-function is normalizable in the same Stoke
wedges as in the case of $V_{L_1}$ in Eq. (\ref{coup}). Thus, the
Hamiltonian $-\hat{H}_{L_2}$ has a well-defined ground-state with
normalizable wave-function. Further, the box-normalization may be used
to obtain complete spectra.

\subsection{Rational CMS-type many-body systems}

The BO describing a balanced loss and gain system in a harmonic well and
without any coupling does not allow classical stable solutions. 
However, stable classical solutions can be obtained for various choices 
of the potential $V$ in (\ref{Ham}) \cite{ben,ben1,wgm,ds-pkg, pkg-ds}. 
In Ref. \cite{pkg-ds} a many particle model with four-body interaction in one 
dimension in the presence of balanced loss and gain terms is introduced 
and its exact stable classical solutions are obtained. The purpose of this 
section is to discuss the quantization and to find exact solutions of some 
of the Calogero-type many-body systems in the presence of balanced 
loss and gain. In particular, the quantization of a many-body 
system with balanced loss and gain and are interacting via 
four-body inverse square potential plus pair-wise two-body harmonic term 
is considered. In $z_i^-$ coordinate this potential takes the following form:

\bea
V_I(z_i^-)&=& - \frac{\omega^2}{2} \sum_{i=1}^m (z^-_i)^2-
\sum_{\substack{i,j=1 \\ i < j}}^m\frac{g}{(z^-_i-z^-_j)^2}.
\label{pot}
\eea
Apart from a negative sign, the potential in Eq. (\ref{pot}) is the
reminiscent of rational CMS model describing a one dimensional
many-particle system interacting with each other via a long range 
inverse square potential and are confined by a common harmonic potential.
This potential is exactly solvable and can be mapped to a set of free harmonic
oscillators\cite{gurappa}. The inverse square part of the potential remains
invariant under any constant shift of the coordinates $z_i^-$. It should be
noted that the potential (\ref{pot}) in transformed $z_i^-$ coordinates
admits a permutation symmetry which is partly lost, when it is expressed in
terms of the original coordinates $x_i$:
\bea
V_I&=&\sum_{i=1}^m -\frac{\omega^2}{4} (x_{2i-1}-x_{2i})^2-\sum_{\substack{i,j=1 \\ i < j}}^m\frac{2g}{(x_{2i-1}-x_{2i}-x_{2j-1}+x_{2j})^2}.
\label{vfour}
\eea
\noindent Many-body systems with four-body interaction have been considered
earlier. For example, the exactly solvable quantum models of CMS-type
with translational invariant two and four-body interactions is investigated
in Ref. \cite{hack} and exactly solvable model with translational
non-invariant four-body interaction is discussed in Ref. \cite{bac}. The most
general four-body inverse-square interaction for a many-particle system with
$2m$ number of particles has the form\cite{wolf}:
\be
V_4=\sum_{\substack{i,j,p,q=1 \\ i \ne j\ne p\ne q }}^{2m}\frac{g}
{(x_i-x_j-x_p+x_q)^2}.
\label{V4}
\ee
\noindent This potential admits a permutation symmetry and remains invariant
under all the operations of a permutation group $S_{2m}$. On the other hand,
the potential $V_I$ is not invariant under all the operations of a permutation
group $S_{2m}$. However, if we map each pair of particles $(x_{2i-1}, x_{2i}),
i=1, \dots, m$ to one element of a permutation group $S_m$, then $V_I$
remains invariant under all the operations of the permutation group
$S_{m}$ and obviously this forms a subgroup of the larger group $S_{2m}$. 
It may be recalled that the Hamiltonian formulation of generic many-body
systems with balanced loss and gain is possible, only when the balancing
of loss and gain terms occurs in a pair-wise fashion. A set of pairs 
$(x_{2i-1}, x_{2i}), i=1, \dots, m$ is chosen in this article out of
$m(2m-1)$ number of possible sets of pairs. The union of elements of
$S_m$ corresponding to each such set gives all the elements of $S_{2m}$. 

The potential $V_I$ is singular at $m(m-1)/2$ points. The configuration
space is divided into $\frac{m(m-1)}{2}$ disjoint sectors. The standard
approach is to solve the eigen-value equation in the one of the sectors, say
$z_1^- < z_2^- < \dots < z_m^-$, and then use the permutation symmetry of
the Hamiltonian to extend the solutions to all the sectors of the configuration
space.  The Hamiltonian $\hat{H}_L$ is not invariant under permutation
symmetry of $S_{2m}$. However, the effective Hamiltonian is invariant under
permutation symmetry under the operations of $S_m$, since $z_i^+$ degrees of
freedom decouples completely. Thus, the same approach as in the case of
rational CMS model may be taken for $V_I$ also.

The second CMS-type model considered in this section has the following
potential:
\bea
V_{II}(z_i^-)= - \frac{\omega^2}{2} \sum_{i=1}^m (z^-_i)^2-
\sum_{i=1}^{m-1}\frac{2g}{(z^-_i-z^-_{i+1})^2}+
\sum_{i=2}^{m-1}\frac{2G}{(z^-_{i-1}-z^-_{i})(z^-_{i}-z^-_{i+1})}.
\label{pot1}
\eea
\noindent The potential $V_{II}$ describes short-range nearest-neighbour
and next-to-nearest neighbour interactions that was introduced and studied
in Ref. \cite{jain-khare}. The model is not completely solvable. However,
infinitely many exact states can be obtained analytically. The ground-state
wave-function is related to the joint probability density function of the
eigenvalues of short-range Dyson models\cite{jain-khare}. The potential 
$V_{II}$, when expressed in $x_i$ coordinates, describes short-range four-body
and six-body inverse-square interactions in presence of pair-wise harmonic
confinement.

\subsubsection{Solution for $V_I$}

For the potential $V_I$ in Eq. (\ref{pot}) and with $k_j=k \ \forall \ j$, Eq.
(\ref{iso}) reduces to the following form:
\bea
\sum_{j=1}^m\left(-\partial^2_{z_j}+ \Omega^2 z_j^2\right)\psi+\sum_{\substack{i,j=1 \\ i < j}}^m\frac{g}{\left(z_i-z_j\right)^2}\psi
=-\tilde{E}\psi.
\label{sc}
\eea
\noindent The Eq. (\ref{sc}) takes the form of a many body system interacting
via an inverse-square two-body potential in a harmonic well\cite{sut} with
angular frequency $\Omega$.
The solutions are taken to have the following form:
\bea
\psi&=&\prod_{\substack{i,j=1 \\ i < j}}^m(z_i-z_j)^{\lambda}\phi(r)P_l(z), \ \ \ \ r^2=\sum_{i=1}^m z_i^2,
\ \ \ \lambda=\frac{1}{2}[1+(1+4g)^{\frac{1}{2}}],
\label{psi}
\eea
where $P_{l}(z_1, \dots, z_m)$ is a homogeneous polynomial of degree $l\ge 0$
satisfying the generalized Laplace equation\cite{calo}.
The normalizable solution of $\psi(r)$ has the following form\cite{sut}:
\bea
\phi_{n}(r)&=&\exp{[-\frac{1}{2}\Omega r^2]}L_n^b[\Omega r^2],\ \ \ \ n=0,1,2,3...
\eea
with $L^b_n$ being the Laguerre polynomial and $b=\frac{-\tilde{E}}{2\Omega}-2n-1$.
The energy eigenvalues are\cite{sut}
\bea
E =-2 \Omega [2n+l+\frac{1}{2}m+\frac{\lambda}{2} m (m-1)] +
\frac{m k^2 \omega^2}{2 \Omega^2}.
\eea
The energy eigenvalues become negative and unbounded from below in the $k=0$
sector, unless $\Omega<0$. This in turn necessitates the normalization of the
wave functions in proper Stoke wedges. Since, the asymptotic form of the total
wave function $\chi$ for the potential $V_I$ is given by Eq. (\ref{asymp})
with $k_j=k \forall \ j$, its normalization in proper Stoke wedges is the same
as considered at the end of section-3 for the case of simple harmonic
oscillators. The eigenvalues consists of discrete as well as continuous
spectra. The box-normalization may be used to confine the particles with
continuous spectra within a length $L$ along their co-ordinates. The momentum
vector $k$ gets quantized and the energy $E$ has the expression,
\bea
E =-2 \Omega [2n+l+\frac{1}{2}m+\frac{\lambda}{2} m (m-1)] +
\frac{2m \pi^2 \omega^2 i^2}{ L^2 \Omega^2}, \ \Omega <0, \
i=0, 1,2, \dots.
\eea
\noindent It may be noted that this is not the complete spectra due
to the assumption $k_j = k \forall \ j$. It is not apparent whether or not
analytical eigen spectra may be obtained for $V_I$ with $m$ independent
$k_j$'s.

The eigenvalue problem of $\hat{H}_{L_2}$ for the potential $V_{L_2}(z_i^+)=-V_I(z_i^+)$ can
be studied by using the duality transformation (\ref{duality}). In $x_i$
coordinates the potential $V_{L_2}$ has the following form

\bea
V_{L_2}(x_i)=\sum_{i=1}^m \frac{\omega^2}{4} (x_{2i-1}+x_{2i})^2+
\sum_{\substack{i,j=1 \\ i < j}}^m\frac{g}{2(x_{2i-1}+x_{2i}-x_{2j-1}-x_{2j})^2}.
\eea
 It is evident that although the functional form of $V_{L_2}$ and $V_I(z_i^-)$ are same, they
correspond to entirely different physical systems when expressed in the coordinates $x_i$. 
The energy spectra for  the eigenvalue problem of $\hat{H}_{L_2}$ 
for the potential $V_{L_2}(z_i^+)=-V_I(z_i^+)$ is given by
\bea
E =2 \Omega [2n+l+\frac{1}{2}m+\frac{\lambda}{2} m (m-1)] -
\frac{m k^2 \omega^2}{2 \Omega^2}.
\eea
 The eigenvalue is bounded from above for
$\Omega < 0$ and the wave-function is normalizable in the complex
$z_i^+$ planes in the same Stoke
wedges as in the case of $V_{I}(z_i^-)$ ensuring that the
Hamiltonian $\hat{H}_{L_2}$ has a well-defined ground-state with
normalizable wave-function for the potential $V_{L_2}(z_i^+)$.
In this case the box-normalization may also be used to obtain
complete spectra as discussed for the Hamiltonian $\hat{H}_{L_1}$ 
with the potential $V_I(z_i^-)$.

\subsubsection{Correlation functions for the potential $V_I$}

The ground state wave function for the potential $V_I$ is given by:
\bea
\chi&\sim&\prod_{\substack{i,j=1 \\ i < j}}^m|z_i-z_j|^{\lambda}
\exp \left [-\frac{\Omega}{2} \sum_{j=1}^m z_j^2 \right],
\label{sol}
\eea
and $\chi$ is independent of $z_i^+$. A scale transformation 
$y_i=\sqrt{\frac{\Omega}{\lambda}}z_i$ of $z_i$ coordinates gives
\bea
|\chi|^2&=&C\prod_{\substack{i,j=1 \\ i < j}}^m|y_i-y_j|^{2\lambda}\exp
\left(\sum_{j=1}^m-\lambda y_i^2\right),
\label{xi2}
\eea
where C is the normalization constant and is given by
\bea
C^{-1}=\int_{-\infty}^{\infty}......\int_{-\infty}^{\infty}
{\mid \chi(y_1,...y_m) \mid}^2 \prod_{i=1}^m dy_idz^+_i.
\eea
Since, $\Omega<0$ and $\lambda >0$, the scale transformation
$y_i=\sqrt{\frac{\Omega}{\lambda}} z_i$ also involves a rotation by 
$\frac{\pi}{2}$ in the complex $z_i$ plane which takes care that all the
relevant integrations are carried out in proper Stoke wedge.
It should be noted that the system is not normalizable 
unless it is bounded along the $z_i^+, \ \forall \ i$ directions.
It is assumed that the system is bounded within the length 
$L$ along $z_i^+, \forall \ i$ directions. The expression for
 $C$ may now be written as\cite{sut}:
\bea
C=\frac{[\Gamma(1+\lambda)]^m}{L^m(2\pi)^{\frac{1}{2}m}
(2\lambda)^{-\frac{1}{2}m-\frac{1}{2}\lambda m(m-1)}
\prod_{j=1}^m\Gamma(1+\lambda j)}.
\eea
\noindent The constant $C$ vanishes in the limit $L \rightarrow \infty$,
since the particles in $z_i^+$ directions are no more confined. 
The n-particle correlation function for the model 
$V_I$ in the $x_i$ coordinate may be defined as:
\bea
R_n(x_1, x_2,.....x_n)=\frac{N!}{(N-n)!}\int_{-\infty}^{\infty}......
\int_{-\infty}^{\infty} {\mid \chi(x_1,x_2,....,x_N) \mid}^2 
\prod_{i=n+1}^{N} dx_i, \ n < N.
\label{Rn}
\eea
It should be noted that the coordinate transformation  (\ref{trans}) 
maps the original problem of many-body systems with balanced loss 
and gain and are interacting via four-body inverse square potential 
plus pair-wise two-body harmonic term to the rational CMS model 
with a common harmonic confinement. Therefore, it is expected
 that the correlation functions of Eq. (\ref{Rn}) can be mapped to 
the correlation function corresponding to the rational CMS model 
with a common harmonic confinement. However, due the special 
nature of the transformation (\ref{trans}), only the even $2n$ 
point correlation functions of Eq. (\ref{Rn}) in $x_i$ coordinate 
can be mapped to the rational CMS model. In particular, 
Eq. (\ref{Rn}) may be rewritten as:
\bea
R_{2n}(x_1, x_2,...,x_{2n})=\frac{N!}{(N-2n)!}\int_{-\infty}^{\infty}......
\int_{-\infty}^{\infty} {\mid \chi (x_1,x_2...,x_N) \mid}^2
\prod_{i=n+1}^{m} dx_{2i-1}dx_{2i}, n<m.
\label{Rn1}
\eea
which after the coordinate transformation of the 
form (\ref{trans}) takes the following form:
\bea
R_{2n}=\frac{N!}{(N-2n)!}\int_{-\infty}^{\infty}......
\int_{-\infty}^{\infty} {\mid \chi(z_1^+,...z_m^+,z_1^-,...z_m^-)\mid}^2
\prod_{i=n+1}^{m} dz^+_idz^-_i.
\label{Rn2}
\eea
Since $|\chi|^2$ is independent of $z_i^+$ coordinates, Eq. (\ref{Rn2}),
after integrating over the $z_i^+$ variables takes the following form:
\bea
R_{2n}=\frac{N!L^{m-n}}{(N-2n)!}\int_{-\infty}^{\infty}......
\int_{-\infty}^{\infty} {\mid \chi(z_1^-,...z_m^- \mid}^2
\prod_{i=n+1}^m dz^-_i.
\label{Rn4}
\eea
Substituting the expression for $|\chi|^2$ from 
Eq. (\ref{xi2}), the following Eq. is obtained
\bea
R_{2n}=\frac{C N!L^{m-n}}{(N-2n)!}\int_{-\infty}^{\infty}...
\int_{-\infty}^{\infty}\prod_{\substack{i,j=1 \\ i < j}}^m
|y_i-y_j|^{2\lambda}\exp\left(\sum_{j=1}^m-
\lambda y_i^2\right)\prod_{i=n+1}^m dy_i.
\label{Rn3}
\eea
The right hand side of Eq. (\ref{Rn3}), apart form a multiplicative 
factor, gives the $2n$-particle correlation function for
rational CMS  model with a common harmonic confinement. 
For $n=1$, the $2$-particle correlation function is obtained as:
\bea
R_2 =
\begin{cases}
\frac{N(N-1)}{m\pi L}(2m-y^2)^{\frac{1}{2}}, \ \ y^2<2m\\
0,  \ \ \ \ \ \ \ \ \ \ \ \ \ \ \ \ \ \ \ \ \ \ \ \ y^2>2m.
\end{cases}
\eea
For $n=2$, the $4$-particle correlation function is obtained as:
\bea
R_4=\frac{N!(m-2)!}{L^2m!(N-4)!}(1-Y(s)),\ s=\frac{2\sqrt{m}}{\pi}|y_1-y_2|.
\eea
The expressions of $Y(s)$ for various $\lambda$ are as given in Ref.\cite{sut}.

It should be noted that the correlation functions  for the eigenvalue problem of 
$\hat{H}_{L_2}$ for the potential $V_{L_2}(z_i^+)=-V_I(z_i^+)$ 
can be obtained in a similar manner as discussed above for the eigenvalue problem of 
$\hat{H}_{L_1}$ for the potential  potential $V_{I}(z_i^-)$. This is an
important result due to the duality symmetry (\ref{duality}), since the potentials 
$V_{L_2}(z_i^+)$ and $V_I(z_i^-)$ correspond to entirely different physical systems 
when expressed in the $x_i$ coordinates.

\subsubsection{Solution for $V_{II}$}

For the potential $V_{II}$ in Eq. (\ref{pot1}) and $k_j=k \forall \ j$,
Eq. (\ref{iso}) reduces to the following form:
\bea
&&\sum_{j=1}^m\left(-\partial^2_{z_j}+ \Omega^2 z_j^2\right)\psi
+\sum_{i=1}^{m-1}\frac{2g}{\left(z_i-z_{i+1}\right)^2}\psi
-\sum_{i=2}^{m-1}\frac{2G \psi}{(z_{i-1}-z_{i})(z_{i}-z_{i+1})}=-\tilde{E}\psi,
\label{sc1}
\eea
and the solutions are taken to be of the form:
\bea
\psi_n&=&\prod_{i=1}^{m-1}(z_i-z_{i+1})^{\lambda}\phi(r)P_l(z), 
\ \ \ \ g=\lambda(\lambda-1), \ \ G=\lambda^2.
\label{psi1}
\eea
where $P_l(z_1, \dots, z_m)$ is a translation-invariant homogeneous polynomial of 
degree $l \geq 0$\cite{jain-khare}.
The normalizable solutions for $\phi$ can readily be obtained as\cite{jain-khare}:
\bea
\phi_{n}(r)&=&\exp{[-\frac{1}{2}\Omega r^2]}L_n^b[\Omega r^2],\ \ \ \ n=0,1,2,3...
\eea
with $L^b_n$ being the Laguerre polynomial and $b=\frac{-\tilde{E}}{2\Omega}-2n-1$.
The energy eigenvalues are
\bea
\tilde{E}=-2\Omega [2n+l+\frac{m}{2}+\lambda(m-1)].
\eea
It should be noted that the energy eigenvalues become negative, unless
$\Omega<0$.
This fact necessitates the normalization of the wave functions in proper Stoke
wedges. 
Since, the total wave function $\chi$ corresponding to the potential $V_{II}$ has the 
asymptotic form of Eq. (\ref{asymp}), the normalization in proper Stoke wedges
 is the same as considered at the end of section-3. The expression for $E$
and its fully quantized form may be obtained in a similar way as in the case
of $V_I$ and may be written as

\bea
E=-2\Omega [2n+l+\frac{m}{2}+\lambda(m-1)]+\frac{2m \pi^2 \omega^2 i^2}{ L^2 \Omega^2}, \ \Omega <0, \
i=0, 1,2, \dots.
\eea
\noindent It may be noted that this is not the complete spectra due
to the assumption $k_j = k \forall \ j$. It is not apparent whether or not
analytical eigen spectra may be obtained for $V_I$ with $m$ independent
$k_j$'s.

The eigenvalue problem of $\hat{H}_{L_2}$ for the potential $V_{L_2}(z_i^+)=-V_{II}(z_i^+)$ can
also be studied by using the duality transformation (\ref{duality}). 
The energy spectra for  the eigenvalue problem of $\hat{H}_{L_2}$ 
for the potential $V_{L_2}(z_i^+)=-V_{II}(z_i^+)$ may be written by 
exploiting the duality symmetry, in the following form:
\bea
E =2 \Omega [2n+l+\frac{m}{2}+\lambda(m-1)] -
\frac{m k^2 \omega^2}{2 \Omega^2}.
\eea
 The eigenvalue is bounded from above for
$\Omega < 0$ and the wave-function is normalizable in the same Stoke
wedges in the complex $z_i^+$ planes as in the case of $V_{II}(z_i^-)$.
In this case the box-normalization may also be used to obtain
complete spectra as discussed for the Hamiltonian $\hat{H}_{L_1}$ 
with the potential $V_{II}(z_i^-)$.

\subsubsection{Correlation functions for the potential $V_{II}$}

The ground state wave function for the potential $V_{II}$ is given by:
\bea
\chi&\sim&\prod_{i=1}^{m-1}|z_i-z_{i+1}|^{\lambda}
\exp\left[-\frac{\Omega}{2} \sum_{j=1}^m z_j^2 \right],
\label{solnew}
\eea
\noindent which is independent of $z_i^+$. A scale
transformation $y_i=\sqrt{\frac{\Omega}{\lambda}}z_i$ of $z_i$ coordinates
gives,
\bea
|\chi|^2&=&C\prod_{i=1}^{m-1}|y_i-y_{i+1}|^{2\lambda}\exp
\left(\sum_{j=1}^m-\lambda y_i^2\right),
\label{xi2new}
\eea
where C is the normalization constant. The box normalization and the integration
over $z_i^+$ coordinate are as discussed before for the case of calculating the 
correlation function for the potential $V_I$. Apart from a constant factor
the two  and four particle correlation functions for the potential $V_{II}$
can be mapped to the one and two point correlation functions as discussed in 
Ref. \cite{jain-khare} for the potential $V_{II}$. Therefore, in the present case 
the  two and four particle correlation functions for the potential $V_{II}$
are respectively given as:
\bea
R_2&=&\frac{N(N-1)}{2\pi L}\frac{\sqrt{|\Omega|}}{\sqrt{2\pi(1+\lambda)}}exp[-\frac{|\Omega| z_1^2}{2(1+\lambda)}],\\
R^{\lambda}_4(s)&=&\frac{N!(m-2)!}{L^2m!(N-4)!}\sum_{k=1}^{\lambda}B^k exp[(1+\lambda)s\{B^k-1\}],\ B=exp(\frac{2i\pi}{1+\lambda}),
\eea
for any integer $\lambda$.

It should be noted that the correlation functions  for the eigenvalue problem of 
$\hat{H}_{L_2}$ for the potential $V_{L_2}(z_i^+)=-V_{II}(z_i^+)$ 
can also be obtained in a similar manner as discussed above for the eigenvalue problem of 
$\hat{H}_{L_1}$ for the potential  potential $V_{II}(z_i^-)$. This is an
important result and is the manifestation of the duality symmetry (\ref{duality}).
It should be noted that although the potentials $V_{L_2}(z_i^+)$ and $V_{II}(z_i^-)$ have
the same functional form, they correspond to entirely different physical systems 
when expressed in the $x_i$ coordinates.

\section{Rotationally invariant potential}

The Lagrangian of BO is invariant under hyperbolic rotation and the same
symmetry persists in the quantum case. Stable classical solutions are not
possible, if the two oscillators of Batman Hamiltonian are coupled
maintaining this rotational symmetry and the result remains unchanged for
a generalized model with $2m$ particles\cite{pkg-ds}. The quantum problem
of the Hamiltonian in Eq. (\ref{qm11}) with a rotationally symmetric
potential is considered in this section. The radial variable $R$ in the
pseudo-Euclidean co-ordinate with the metric $g_{ij}=(-1)^{i+1} \delta_{ij}$ is defined as,
\be
R^2=2\sum_{i=1}^{m} x_{2 i-1} x_{2 i} = \sum_{i=1}^m \left [ \left ( z_i^+
\right )^2 - \left ( z_i^- \right )^2 \right ],
\label{N-radi}
\ee
\noindent and the potential is taken to be $V \equiv V(R)$. It should be noted that in 
this case the condition (\ref{condi_p2t}) is automatically satisfied and the Hamiltonian becomes both
${\cal{PT}}$ and ${\cal{P}}_1{\cal{T}}$ symmetric. The Hamiltonian
in Eq.  (\ref{qm11}) with $\gamma=0$ is rotational invariant. It may be
noted that the rotational invariance of the Hamiltonian is partially lost
for $\gamma \neq 0$, since the term linear in $\gamma$ is the sum of angular
momenta for rotations in $m$ planes specified by `$z_i^--z_i^+$'. The
Hamiltonian is invariant under rotation when the planes of
rotations are chosen as `$z_i^--z_i^+$'.
The $m$  integrals of motion $L_i$
may be defined for this system which are in involution. In particular,
\be
L_i = -i \left (z_i^+ \frac{\partial}{\partial z_i^-} +
z_i^- \frac{\partial}{\partial z_i^+} \right ), \ \
\left [ L_i, L_j \right ] = 0, \ \ \left [ H, L_i \right ] = 0,
\ee 
\noindent implying that the system is at least partially integrable. It may
be noted that the integrals of motion can also be identified as generators
of rotation in $m$ planes of rotations `$z_i^--z_i^+$'.

It is found in Ref. \cite{pkg-ds} that the classical solutions for a 
chain of nonlinear oscillators belonging to this class of models are unstable. 
It may be shown that the quantum Hamiltonian with generic rotationally
invariant potential $V(R)$ does not admit entirely real spectra. There are
complex as well as real eigenvalues in the spectra, implying the existence of
growing and decaying modes. To see this, an imaginary scale
transformation,
\be
z_i^-\rightarrow -iz^-_i, \ \ P_{z_i^-} \rightarrow i P_{z^-_i}, \ \
z_i^+\rightarrow z_i^+, \ \ P_{z_i^+} \rightarrow P_{z_i^+},
\ee
\noindent is performed on the Hamiltonian (\ref{quantumH}) which reduces
to the following form:
\bea
H & = & {\cal{H}} + i \gamma {\cal{L}},\nonumber \\
{\cal{H}} & = & \sum_{i=1}^m ( P_{z_i^+}^2 +P_{z_i^-}^2) 
-\frac{\gamma^2}{4} {\tilde{R}}^2 + V(\tilde{R}), \ \
{\cal{L}} = \sum_{i=1}^m \left ( z_i^+ P_{z_i^-} - z_i^- P_{z_i^+} \right ),
\eea
\noindent where $\tilde{R}$ is the radial variable in the $2m$ dimensional
Euclidean space. It may be noted that after the imaginary scale transformation
has been performed, the quantum problem is defined in a Hilbert space in which
${\cal{H}}$ and ${\cal{L}}$ are hermitian\cite{ben-man, ali}.  The operator
${\cal{L}}$ has the interpretation of the
sum of angular momentum operators in Euclidean space corresponding to
rotations in $m$ planes of rotations `$z_i^--z_i^+$'.
The operators  ${\cal{H}}$ and ${\cal{L}}$ commute with each other and
therefore admit simultaneous eigen states. With the suitable choices of
$V(\tilde{R})$, entirely real spectra with normalizable eigenfunctions
may be obtained for ${\cal{H}}$. However, due to the presence of the imaginary
coupling $i\gamma$, the eigenvalues of $H$ become complex with the
exceptions for the cases corresponding to the zero eigenvalue of the operator
$ {\cal{L}}$. As an example, the Hamiltonian\cite{ben-man,ali} of the
Pais-Uhlenbeck oscillator(PUO) after an imaginary scale transformation of
variables as above has been performed is similar to ${\cal{H}}$ for the choice
of the potential,
\be
V(\tilde{R})=\frac{\omega^2}{4} \tilde{R}^2, \ \omega^2 > \gamma^2. 
\ee
\noindent It is known\cite{ben-man,ali} that consistent quantum mechanical
description with well defined bound states is possible for PUO. Thus, the
Hamiltonian $H$ does not admit entirely real spectra due to the coupling
$i \gamma$ in the second term. It follows from the
discussions above that the necessary condition for a
Hamiltonian describing a system with balanced loss and gain to admit entirely
real spectra is,
\be
\left [ V(\{z_i^-\}, \{z_i^+\}) , L_i \right ] \neq 0 \ \forall \ i.
\ee
\noindent In other words, the vanishing commutators among 
$V(\{z_i^-\}, \{z_i^+\})$ and $L_i$'s  imply that the spectra contains
complex eigenvalues. Consequently, no bound states can be formed.

\section{Systems with $O(2,1)$ symmetry:}

The BO has dynamical $SU(1,1)$ symmetry\cite{blasone} which can be used to
find its complete spectra. The purpose of this section is to investigate
systems with balanced loss and gain with specific class of $V$ such that
dynamical $SU(1,1)$ is preserved. Further,  it is known that systems with
inverse-square interactions are examples of conformal quantum mechanics and
appear in diverse branches of physics\cite{dff,pkg1,pkg3}. The basic underlying
symmetry of such systems are $O(2,1)$ which is related to $SU(1,1)$. The
rational CMS models belongs to this class and the entire spectra corresponding
to radial excitations may be found by using the underlying $O(2,1)$ symmetry.
The CMS-type models discussed in Sec. 3.1.1 appear as an effective system with
reduced degrees of freedom. In this section, $O(2,1)$ symmetric systems with
$N=2m$ degrees of freedom are discussed.

Three operators $H_0$, $D$ and $K$ are defined in the following fashion:
\bea
H_0&=&\frac{1}{2} \sum_{i=1}^m \left(\Pi^2_{z_i^+}-
\Pi^2_{z_i^-}\right)+V_c(z^-_i,z^+_i),\nonumber\\
D&=&-\frac{1}{4}\sum_{i=1}^m \left(z_i^+\Pi_{z_i^+}
+z_i^-\Pi_{z_i^-}+\Pi_{z_i^+}z_i^++\Pi_{z_i^-}z_i^-\right),\nonumber\\
K&=&\frac{1}{2}\sum_{i=1}^m\left({z_i^+}^2-{z_i^-}^2\right)= \frac{R^2}{2},
\eea
\noindent where the potential $V_c(z^-_i,z^+_i)$ is a homogeneous polynomial 
of degree $-2$,
\be
\sum_{i=1}^m \left [ z_i^- \partial_{z_i^-} + z_i^+ \partial_{z_i^+} \right ]
V_c(z^-_i,z^+_i) = - 2 V_c(z^-_i,z^+_i).
\ee
\noindent The Hamiltonian is ${\cal{PT}}$ symmetric for $V_c(z_i^-,z_i^+)=
V_c(z_i^-,-z_i^+)$. The operator $K$ is ${\cal{PT}}$ symmetric, while $D$
is not invariant under ${\cal{PT}}$. The operators $H_0, D$ and $K$ satisfy
the $O(2,1)$ algebra:
\bea
[H_0,D]=iH,\ \ \ \ [H_0,K]=2iD,\ \ \ \ \ [D,K]=iK.
\eea
\noindent The operators, 
\bea
K_1=\frac{1}{2}\left(K+H_0\right), 
\ \ \ \  K_2=\frac{1}{2}\left(K-H_0\right),\ \ \ \  K_3=D,
\eea
satisfy the $SU(1,1)$ algebra:
\bea
[K_1,K_2]=i K_3, \ \ \ \  [K_1,K_3]=i K_2,\ \ \ \ [K_3,K_2]=i K_1.
\eea
\noindent With the choice of the potential $V(z_i^-,z_i^+)$ in Eq.
(\ref{quantumH}) as,
\be
V(z_i^-,z_i^+)= 2V_c(z_i^-,z_i^+) + \beta^2 R^2,
\ee
\noindent the Hamiltonian $H$ can be expressed in terms of $H_0$ and $K$
as,
\be
H=2(H_0 + \beta^2 K).
\ee
\noindent Note that $H$ is ${\cal{PT}}$ symmetric.
In the limit of vanishing $\beta$, $H$ and $H_0$ are
related as $H=2H_0$. It is argued in the previous section that the spectra is
not entirely real for rotationally invariant $V$. In order to avoid such
situation, it is assumed that $V_c$ is not rotationally invariant. 

The eigenvalue problem of $H$ is discussed below with the help of the
$O(2,1)$ symmetry. Following coordinate transformation is employed:
\bea
z_1^+&=&R \cosh{\theta_1},\nonumber\\
z_1^-&=&R \sinh{\theta_1}\cosh{\theta_2},\nonumber\\
&&......................................\nonumber\\
z_m^+&=&R\sinh{\theta_1}\sinh{\theta_2}\sinh{\theta_3}....... \cosh{\theta_{2m-1}},\nonumber\\
z_m^-&=&R\sinh{\theta_1}\sinh{\theta_2}\sinh{\theta_3}....... \sinh{\theta_{2m-1}}.
\label{hypc}
\eea
In this $2m$ dimensional coordinate, the Hamiltonian $H$ can be re-written
as, 
\bea
H=\Pi^2_R+\beta^2R^2+\frac{\tilde{C}}{R^2},
\ \ \ \ \tilde{C}=4C-m\left(m-2\right),
\eea
\noindent where $\Pi^2_R=-(\frac{\partial^2}{\partial R^2}+
\frac{(2m-1)}{R}\frac{\partial}{\partial R})$ is the radial part of the
conjugate momenta, $C$ is the Casimir operator of the $O(2,1)$ symmetry,
\bea
C&=&\frac{1}{2}\left(H_0K+KH_0\right)-D^2\\
&=&\frac{1}{4}\left[\sum_{i<j}^m -L^2_{ij}+R^2V+m(m-2))\right],
\eea
\noindent and $L_{ij}=\left(z_i^+\Pi_{z_j^-}+z_i^-\Pi_{z_j^+}\right)$ are
the components of angular momentum operators. 
The Hamiltonian $H$ can always be separated into a radial
part and an angular part in the $2m$ dimensional coordinate defined
by Eq. (\ref{hypc}), since the term $R^2V$ contains only the angular variables.
In particular, the equation for the radial variable $R$ may be obtained
from the time independent Schrodinger equation
$H \xi(R) Y(\theta_i, \phi_i) =E \xi(R) Y(\theta_i, \phi_i)$ 
in the following form,
\bea
\frac{d^2\xi}{dR^2}+\frac{(2m-1)}{R}\frac{d\xi}{dR}+(E-\beta^2 R^2-
\frac{{C}^{\prime}}{R^2})\xi=0,
\eea
\noindent where $C^{\prime}$ is the eigenvalue of the operator $\tilde{C}$.
The normalizable solution of the radial variable equation can readily be
obtained as:
\bea
\xi(R)=C_N R^{2s}exp{[-\frac{\beta}{2}R^2}]
L_n^{2s+m-1}(\beta R^2),\ \ \ \ \ s=\frac{-(m-1)+\sqrt{C^{'}+(m-1)^2}}{2},
\eea
\noindent where $C_N$ is normalization constant. The energy spectrum is given
by:
\bea
E_n=4\beta(n+s+\frac{m}{2}).
\eea
The asymptotic form of the wave-function may be written as:
\bea
\xi(R)\sim exp{[-\frac{\beta}{2}\sum_{i=1}^m({(z^+_i})^2-({z_i^-})^2)}].
\label{asy}
\eea
\noindent The 2nd term in the exponential of Eq. (\ref{asy}) for $\beta >0$,
i. e. $ \frac{\beta}{2}\sum^m_{j=1}({z^-_j})^2$, is a source of divergence.
However, $\sum_{j=1}^m ({z^-_j})^2$ vanishes in a pair of Stoke wedges with 
opening angle $\frac{\pi}{2}$ and centered about the positive
and negative imaginary axes in the complex $z^-_j$-planes. It may be noted
that a complete knowledge of the spectra involves analysis of eigenvalue
equation for the angular variables. The form of the potential is to be
specified for such an analysis which is left for future investigation.

\section{Summary and discussion}

The quantization of many-body systems with balanced loss and gain of Ref.
\cite{pkg-ds} has been investigated in this article. It has been argued that
the quantum Hamiltonian can be interpreted as an interacting many-body system
in the background of a pseudo-Euclidean metric and subjected to uniform
``magnetic field" proportional to the gain/loss parameter $\gamma$. The
analogous ``magnetic field" is perpendicular to each plane formed by the pair
of co-ordinates related to balanced loss and gain. Further, it is shown that
either symmetric gauge or Landau gauge may be used to solve the quantum
problem. The Hamiltonian corresponding to these two gauge are related through
an unitary transformation and at the classical level, the corresponding
Lagrangian differ by a total time derivative term. 

Two types of many-body systems characterized by either (i) translational
invariance or (ii) rotational invariance in a space endowed with the metric
$g_{ij}=(-1)^{i+1}\delta_{ij}$ have been considered. For the case of translational invariant
systems,  the Landau gauge Hamiltonian has been used, since the wave-function
of a symmetric gauge Hamiltonian is not suitable for box-normalization for the
co-ordinates associated with continuous spectra. This is a reminiscent of the
fact that two Lagrangian differing by a total time-derivative term may not
lead to the same quantum theory, although the classical dynamics is identical
for the two cases. Thus, the correct route for quantization of the system is
to start from the Lagrangian that leads to Landau gauge Hamiltonian, instead of
obtaining $\hat{H}_{L_1}(H_{L_2})$ from $\hat{H}$ via the unitary transformation.

For translational invariant system, the original eigenvalue equation in terms
of $2m$ degrees of freedom is reduced to an eigen-value equation of an
effective Hamiltonian
in terms of $m$ degrees freedom. This reduction is possible due to the
existence of $m$ integrals of motion. If the eigen-value problem for this
effective Hamiltonian is solvable, then the starting Hamiltonian is also
solvable.
Three examples have been considered: (i) coupled harmonic oscillators,
(ii) rational CMS-type many-body systems with balanced loss and gain, where
each particle is interacting with other particles via four-body inverse-square
potential plus pair-wise two-body harmonic terms and (iii) a many-body system
interacting via short-range four-body plus six-body inverse square potential
with pair-wise two-body harmonic terms. The eigenvalues of these systems are
partly discrete and partly continuous. The box-normalization has been used
for the co-ordinates associated with continuous spectra and a fully quantized
eigen spectra have been obtained. Further, proper Stoke's wedge has been 
identified so that energy is bounded from below and the corresponding
eigenfunctions are normalizable. The exact correlation functions have been
obtained for models (ii) and (iii). Apart from an overall multiplication factor,
these correlation functions are identical with the corresponding quantities
for the effective Hamiltonian.

A partial set of integrals of motion has been obtained for many-body systems
with balanced loss and gain and generic rotationally invariant potential. It has
been argued on general ground that eigen spectra of such systems are not
entirely real and unstable quantum modes are present. Radial excitations are
obtained analytically for systems with $O(2,1)$ symmetry.

Examples of exactly 
solvable models are very rare to find in physics. The models considered 
in this article include many-body systems with balanced loss and gain
for which exact eigenvalues, eigenfunctions and more importantly, a 
few-particle correlation functions are obtained analytically. It appears that no other exactly solvable many-body system with balanced loss and gain having nonlinear coupling is known in the literature for three or more particles. Thus, the model 
presented being only one of its kind, the result is immensely significant from the 
viewpoint of exactly solvable and integrable models.

A very pertinent question one would like to pose is whether or not the solvable
model is of sufficient interest from the viewpoint of physical applications?
It is not apparent whether there exists any specific physical set-up, where the
proposed system can be realized experimentally. However, the importance of the
results lies elsewhere. It may be recalled that the Calogero model does not
directly represent any experimentally realizable system, yet, it appears in the
study of various diverse subjects, ranging from cosmology to condensed matter
systems. Its relevance in the context of exclusion statistics \cite{exst},
symmetric polynomials \cite{c1}, random matrix theory \cite{AD,SJ}, Yang-Mills
theory \cite{pkgc}, conformal quantum mechanics \cite{pkgc,c0}, spin
chains \cite{hal} with long range correlations are worth mentioning. An
important aspect that is central to all types of Calogero model is that the
many-body interaction scales inverse-squarely. The models presented in the
article share this property, raising the expectation that some of the
features of the standard Calogero  model may also be present for the systems
under considerations. The example of conformal quantum mechanics is alreday
presented in Sec. 5 within the context of many-body system with balanced loss
and gain. The permutation symmetry in the models described
in Sec. 3 is realized in a restricted form which may lead to a different
kind of statistics obeyed by the particles that should lead to the standard
exclusion statistics in the limit of vanishing loss and gain. Further,
the $n$-particle correlation functions are computed using known results of
random matrix theory, which indicates that the model may be obtained as a
reduction of some matrix models and Yang-Mills theory. A possible connection
of the system with quantum chaotic system is worth exploring.
Within this background, the results of the
article should be seen as a first step towards unraveling a new class
of solvable models with balanced loss and gain which may have relevance
in the wider context.

\section{Acknowledgements}
The work of PKG is supported by grants({\bf SERB Ref. Nos.
SR/S2/HEP-24/2012 \& MTR/2018/001036}) from Science \& Engineering Research
Board(SERB), Department of Science \& Technology(DST), Govt. of India.
{\bf DS} acknowledges a research fellowship from CSIR.


\begin{thebibliography}{10}

\bibitem{bat} H. Bateman, Phys. Rev. {\bf 38}, 815 (1931).

\bibitem{fes} H. Feshbach and Y. Tikochinsky, in A Festschrift for
     I. I. Rabi, Trans. New York. Acad. Sci., Series 2 38, 44
     (1977).


\bibitem{bopp} F. Bopp, Sitz.-Bcr. Bayer. Akad. Wiss. Math.-naturw.
     KI. {\bf67}, (1973).

\bibitem{trikochinsky} Y. Tikochinsky, J. Math. Phys. {\bf19}, 888 (1978).

\bibitem{dekker} H. Dekker, Phys. Rep. {\bf80}, 1 (1981).

\bibitem{rasetti} E. Celeghini, M. Rasetti, and G. Vitiello, Ann. Phys.
     (N.Y) {\bf215}, 156 (1992).

\bibitem{rabin} R. Banerjee and P. Mukherjee, J. Phys. A: Math.
     Gen. {\bf35}, 5591 (2002).

\bibitem{jur} D. Chruscinski and J. Jurkowski, Ann. Phys. (N.Y.) {\bf321},
     854 (2006).

\bibitem{ben} C. M. Bender, M. Gianfreda, S. K. Ozdemir, B. Peng, and
L. Yang, Phys. Rev. A {\bf88}, 062111 (2013).

\bibitem{ben1} C. M. Bender, M. Gianfreda and S. P. Klevansky, Phys. Rev A{\bf90}, 022114 (2014).

\bibitem{ds-pkg} D. Sinha, P. K. Ghosh, Eur. Phys. J. Plus. {\bf132}, 460 (2017);  arXiv:1705:03426.

\bibitem{pkg-ds} P. K. Ghosh, D. Sinha, Ann. Phys. {\bf 388}, 276–304 (2018);  arxive:1707.01122.

\bibitem{ds-pkg6} D. Sinha, P. K. Ghosh, Ann. Phys. {\bf 400}, 109-127 (2019).

\bibitem{pkg6} P. K. Ghosh, Arxive: 1810.04137. 
 

\bibitem{wgm}B. Peng, S. K. Ozdemir, F. Lei, F. Monifi, M. Gianfreda,
G. L. Long, S. Fan, F. Nori, C. M. Bender, and L. Yang,
 Nature Physics, {\bf10}, 394 (2014).

\bibitem{sagar} T. Shah, R. Chattopadhyay, K. Vaidya, S. Chakraborty, Phys. Rev. E {\bf92}, 062927 (2015).




\bibitem{calo} F. Calogero, Jour. Math. Phys. {\bf10}, 2191 (1969),
F. Calogero, Jour. Math. Phys. {\bf10}, 2197 (1969),
 F. Calogero, Jour. Math. Phys. {\bf12}, 419 (1971).
 
\bibitem{sut}  B. Sutherland, J. Math. Phys.(N.Y.){\bf12}, 246 (1971); {\bf12}, 251 (1971); Phys.Rev. A
{\bf4}, 2019 (1971).

\bibitem{ob} M. A. Olshanetsky and A. M. Perelomov, Phys. Rep. {\bf71}, 314 (1981); {\bf94},
6(1983).

\bibitem{poly}A. P. Polychronakos, Phys. Rev. Lett. {\bf69}, 703 (1992);

\bibitem{pkg1} P. K. Ghosh, J. Phys. A: Math. Theor. {\bf45}, 183001 (2012). 

\bibitem{exst} M. V. N. Murthy and R. Shankar, Phys. Rev. Lett. {\bf73}, 3331(1994).


\bibitem{quch}B. D. Simons, P. A. Lee and B. Altshuler, Phys. Rev. Lett.
{\bf72}, 64(1994); S. Jain, Mod. Phys. Lett.  A {\bf 11}, 1201(1996).

\bibitem{hal}F. D. M. Haldane, Phys. Rev. Lett. {\bf60}, 635 (1988); B. S. Shastry, Phys.
Rev. Lett. {\bf60}, 639 (1988).

\bibitem{integ}K. Hikami and M. Wadati, Phys. Rev. Lett. {\bf73}, 1191(19994); H. Ujino and
M. Wadati, J. Phys. Soc. Jap. {\bf63}, 3585(1994).

\bibitem{pkg2}B. Basu-Mallick, P. K. Ghosh and Kumar S. Gupta, Phys. Lett. A {\bf311},
87(2003), hep-th/0208132; B. Basu-Mallick, P. K. Ghosh and Kumar S.
Gupta, Nucl. Phys. B {\bf 659}, 437 (2003), hep-th/0207040; B. Basu-Mallick,
P. K. Ghosh and Kumar S. Gupta, Pramana-J. Phys. {\bf62}, 691 (2004);
B. Basu-Mallick and Kumar S. Gupta, Phys. Lett. A {\bf292}, 36 (2001),
hep-th/0109022.

\bibitem{cfmp} V. Bardek, J. Feinberg, S. Meljanac, JHEP {\bf08}, 018(2010); V. Bardek, J.
Feinberg, S. Meljanac, Annals of Physics {\bf325}, 691 (2010).


\bibitem{qhe}H. Azuma and S. Iso, Phys. Lett. B {\bf331}, 107(1994).

\bibitem{tl} N. Kawakami and S.-K. Yang, Phys. Rev. Lett. {\bf67}, 2493(1991).



\bibitem{wolf} Wolfes, Ann. Phys. {\bf85}, 454 (1974).

\bibitem{hack} O. Haschke, W. Ruhl, arXiv:hep-th/9807194.

\bibitem{bac} A. Bachkhaznadji, M. Lassaut, Few-Body Syst. {\bf54}, 1945 (2013).

\bibitem{blasone} M. Blasone, E. Graziano, O. K. Pashaev, G. Vitiello,
Annals Phys. {\bf 252}, 115 (1996).

\bibitem{diehl} L. M. Sieberer, M. Buchhold, S. Diehl,
Rep. Prog. Phys. {\bf 79}, 096001 (2016), arxiv:1512.00637.


\bibitem{diehl1} L. M. Sieberer, A. Chiocchetta, A. Gambassi, U. C. Täuber,
S. Diehl,
Phys. Rev. B {\bf 92},134307 (2015), arXiv:1505.00912;
L. M. Sieberer, S. D. Huber, E. Altman, S. Diehl,
Phys. Rev. B {\bf 89}, 134310 (2014).
\bibitem{jsm} C. Aron, G. Biroli and L. F. Cugliandolo, J. Stat. Mech. 1011,
P11018, (2010), arXiv:1007.5059.
\bibitem{leewuen} R. V. Leeuwen, N. E. Dahlen, G. Stefanucci,
C-O. Almbladh, U. V. Barth,
arXiv:cond-mat/0506130.
\bibitem{mr} F. M. Haehl, R. Loganayagam and M. Rangamani,
J. High. Energ. Phys., 2017:69 (2017),
arXiv:1610.01940; F. M. Haehl, R. Loganayagam, and M. Rangamani,
J. High. Energ. Phys., 2017:70 (2017),
arXiv:1610.01941.
\bibitem{abc} C. Aron, G. Biroli, L. F. Cugliandolo,
arXiv:1705:10800. 

\bibitem{F.Haake} F. Haake, Quantum Signatures of Chaos, Springer Series in Synergetics(1st edition, 1991).

\bibitem{gurappa} N. Gurappa and P. K. Panigrahi, Phys. Rev. B {\bf59}, R2490(R) (1999).

\bibitem{jain-khare} 
S. R. Jain and A. Khare, Phys. Lett. A {\bf 262}, 35 (1999),
G. Auberson, S. R. Jain, A. Khare, Phys. Lett. A {\bf 267}, 293 (2000);
J. Phys.  A {\bf34}, 695 (2001). 

\bibitem{ben-man} C. M. Bender and P. D. Mannheim, Phys. Rev. Lett. {\bf100}, 110402 (2008).

\bibitem{ali} A. Mostafazadeh Phys. Rev. D {\bf84}, 105018 (2011).

\bibitem{dff}  V. de Alfaro, S. Fubini and G. Furlan, Nuvo Cimento
A{\bf34}, 569 (1976).

\bibitem{pkg3} P. K. Ghosh, J. Phys. A: Math. Gen. {\bf 34}, 5583 (2001). 

\bibitem{c1} P. Desrosiers, L. Lapointe and P. Mathieu, Nucl. Phys. B {\bf606}, 547(2001),
hep-th/0103178.

\bibitem{AD}A. Dabholkar, Nucl. Phys. B {\bf368}, 293(1992).

\bibitem{SJ} S. James Gates Jr., A. Jellal, E. L. Hassan Saidi and M. Schreiber, JHEP
{\bf 0411}, 075(2004).

\bibitem{pkgc} P. K. Ghosh, J. Phys. A {\bf 34}, 5583 (2001);hep-th/0009055.


\bibitem{c0} N. Wyllard, J. Math.Phys. {\bf41}, 2826(2000);  hep-th/9910160.











\end{thebibliography}
\end{document}